\documentclass[useAMS,usenatbib]{mn2e}
\usepackage{amsmath}                                                            
\usepackage{amsfonts}
\usepackage{graphicx}
\usepackage{float}
\usepackage{morefloats}
\usepackage{listings}
\usepackage{todonotes}
\usepackage{color,hyperref}
\definecolor{linkcolor}{rgb}{0,0,0.25}
\hypersetup{
  colorlinks=true,        
  linkcolor=linkcolor,    
  citecolor=linkcolor,    
  filecolor=linkcolor,    
  urlcolor=linkcolor      
}

\usepackage{cleveref}
\usepackage{multicol}
\usepackage{placeins}
\usepackage{pdfpages}
\usepackage[normalem]{ulem}

\title[Tidal Tails of NGC 288]
{Rediscovering the Tidal Tails of NGC 288 with Gaia DR2}
\author[S. Kaderali et al.]
  {\parbox{\textwidth}{Shaziana Kaderali$^{1,2}$\thanks{E-mail: ShazianaKaderali@gmail.com}, Jason A. S. Hunt$^{2}$, Jeremy J. Webb$^{3}$, Natalie Price-Jones$^{3,2}$ and Raymond Carlberg$^{3}$}\vspace{0.5cm}
\\
$^{1}$ Department of Mechanical and Aerospace Engineering, Carleton University, 1125 Colonel By Drive, Ottawa, ON, K1S 5B6, Canada\\
$^{2}$ Dunlap Institute for Astronomy and Astrophysics, University of Toronto, 50 St. George Street, Toronto, ON, M5S 3H4, Canada\\
$^{3}$ Department of Astronomy and Astrophysics, University of Toronto, 50 St. George Street, Toronto, ON, M5S 3H4, Canada \\
}
\date{Accepted 2019 February 1. Received 2019 January 29; in original form 2018 September 10} 

\pagerange{\pageref{firstpage}--\pageref{lastpage}}
\pubyear{2018}

\begin{document}

\maketitle

\label{firstpage}

\begin{abstract}
NGC 288 is a Galactic globular cluster having observed extra tidal structure, without confirmed tidal tails. $Gaia$ DR2 provides photometric and astrometric data for many of the stars in NGC 288 and its extra tidal structure. To compare with $Gaia$ data, we simulate an $N$-body model of a star cluster with the same orbit as NGC 288 in a Milky Way potential. The simulation shows that the cluster forms tidal tails that are compressed along the cluster's orbit when it is at apocentre and are expected to be a diffuse bipolar structure. In this letter, we present a comparison between the simulation and observations from $Gaia$ DR2. We find that both the simulation and the observations share comparable trends in the position on the sky and proper motions of the extra-tidal stars, supporting the presence of tidal tails around NGC 288. 
\end{abstract}


\section{Introduction}
\label{intro}
Globular clusters (GCs) are gravitationally bound groupings of old stars, formed in the early Galaxy. If a cluster is tidally filling \citep[e.g.][]{Henon61}, such that stellar orbits reach the tidal radius (Jacobi radius, $r_{\text{t}}$), we expect to see a signature of stars escaping the cluster via tidal stripping. Tidal stripping results in escaping stars populating tidal tails around the cluster which provide information about its orbit, allowing us to constrain the gravitational potential of the Galaxy \citep[e.g.][]{Bovy_MWstreams}. GCs that are considered to be tidally filling but are lacking tidal tails have likely undergone additional interactions (i.e. tidal shocks) that prevent the formation of tidal tails \citep[e.g.][]{tidal_shock1,DM_288,NGC288}. Alternatively, a high percentage of dark matter can cause an increase in the tidal radius, inhibiting stars from escaping and forming tidal tails \citep[e.g.][]{DM_288,DMNGC288,DMGCs}.

Another potential explanation for a tidally filling GC that is not exhibiting clear tidal tails is simply that they are difficult to observe owing to the current orbital phase of the cluster, or the projection onto the sky of its tail stars, or a combination of the two. With proper motions ($\mu_{\alpha},\mu_{\delta}$) from the second data release \citep[DR2;][]{DR2} of the European Space Agency's $Gaia$ mission \citep{GaiaMission}, we can better constrain the orbits of GCs and determine whether their orbital phase or projection effects affect the detection of tidal tails. For example, when a cluster is at apocentre the tails will contract towards the cluster, reducing their length and visibility. However, the signature of the tails should remain visible in the proper motions of individual stars.

NGC 288 is thought to be a tidally filling GC with no confirmed tidal tails \citep{DM_288}. NGC 288 resides sufficiently nearby \citep[$d = 8.9$ kpc from our Sun;][2010 edition]{GC_cat} that a relatively large number of cluster member stars are within the $Gaia$ limiting magnitude, making it an ideal candidate for further study.

\begin{figure*}
	\centering
    \includegraphics[width=\hsize]{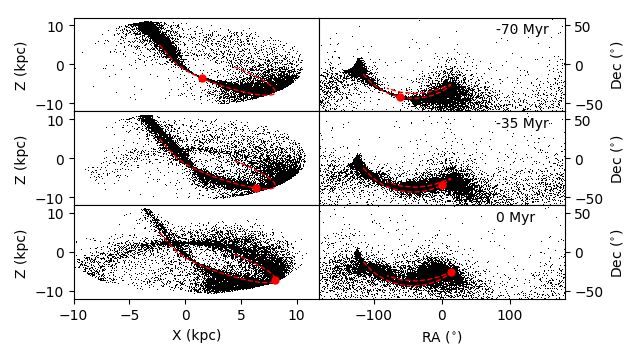}
    \caption{Positions of model stars (black) and cluster centre (red) in both galactocentric coordinates (left column) and equitorial coordinates (right column) between 70 Myr before the cluster reaches apocentre (top row) to its present day position at apocentre (bottom row). The red dashed line shows the orbital path of the model cluster. Tails that were distinct in the top panels, 70 Myr prior to apocentre, have become indistinct by the time the cluster reaches apocentre, especially in RA-Dec.}
    \label{xz_radec}
\end{figure*}

While the lack of observed tidal tails has led to suggestions that NGC 288 contains a large amount of dark matter, it is possible that the non-detection is due to projection effects combined with the fact that the cluster is near or at apocentre \citep{dinescu97,DR2GC}. Additionally, the tidal shock that NGC 288 experienced when passing through the disc may have decreased the extent of the tidal tails \citep{tidal_shock1}. Thus it is expected that any existing tidal tails around NGC 288 would be much shorter than in other clusters with pronounced tidal tails, such as Pal 5 \citep[e.g.][]{Pal5_1,Pal5_2}. 

More recently, \citet{NGC288} and \citet{DM_Shipp} made contrasting observations of extra tidal structure around NGC 288. \citet{NGC288} observes a diffuse halo structure up to $\sim2.5$ times the tidal radius of NGC 288 using PanSTARRS-PS1 \citep{PanSTARRS-1}, while \citet{DM_Shipp} observes extra-tidal structure resembling tidal tails that extends to $\sim 5.5^\circ$ to the south of the GC using the Dark Energy Survey \citep[DES; e.g.][]{DES2}. \citet{DM_Shipp} construct a model of NGC 288 displaying tidal tails, although the orientation of the structure found in their simulation does not match the observation.

In this letter, we explore the possibility that the tidal tails of NGC 288 have been compressed at apocentre and are lost in projection. We make use of proper motion data from $Gaia$ DR2 to explore the extra-tidal structure around NGC 288 and compare to a simulation of a cluster with the same orbit. In \Cref{sim} we present details on the simulation, and in \Cref{obs} we discuss the selection and treatment of data from $Gaia$ DR2. In \Cref{Comp} the simulation is compared to the observations from $Gaia$, while \Cref{Summary} describes our conclusions and the potential for further analyses. 

\section{Simulation}
\label{sim}
To determine how we expect positions and velocities of tidal tail stars of NGC 288 are oriented on the plane of the sky at different phases of its orbit, we perform an $N$-body simulation of a star cluster over the last 4 Gyr of its orbit. The simulation was performed using the \sc{gyrfalcon }\rm code \citep{dehnen00,dehnen02} and a softening length of 1.5 pc in the \sc{nemo }\rm toolkit \citep{teuben95}. The cluster's orbit was integrated using \sc{galpy }\rm \citep{B15} in a Galactic potential assumed to be the MWPotential2014 model from \citet{B15}, with the proper motions, distance and radial velocity of NGC 288 from Gaia DR2 \citep{DR2GC}, which places it very close to apocentre. Stellar positions and velocities for the model cluster were assigned based on a Plummer distribution function with a $37,200 M_{\odot}$ initial mass and a 5 pc scale radius. It should be noted that the initial mass and size of our model cluster are not intended to perfectly reproduce an NGC 288-like cluster after 4 Gyr of evolution, as the main purpose of the model is to determine the properties of stars which have escaped the cluster. 

As the model cluster evolves, stars escape NGC 288 via tidal stripping and populate its tidal tails as expected for a tidally filling GC (see \Cref{xz_radec}). Tidal shocks and the strong tidal field experienced by NGC 288 along its orbit prevent the tails from becoming long and distinct (as otherwise seen in GCs such as Pal 5; e.g. \citealt{Pal5_1,Pal5_2}). However, within a few tidal radii of the model cluster, a clear tidal tail signature is present. The tidal tails are most difficult to observe when the cluster is near apocentre, as the cluster is able to catch up to the leading tail while approaching apocentre, and the trailing tail is able to catch up to the cluster. This behaviour is clear in the left panels of \Cref{xz_radec}, which illustrates the model cluster approaching its current position at apocentre in galactocentric coordinates, where stars to the right of the cluster centre are leading tail members and stars to the left of the cluster centre are trailing tail members. The red dashed line shows the model cluster's orbital path. Observations of the tails become more difficult when the cluster is projected onto the plane of the sky, as seen in the right panels of \Cref{xz_radec}, with projection effectively removing any trace of the tails in position space. For visual purposes, the RA $\geq$ 180$^\circ$ section has been flipped to negative to better illustrate the tails in the position space projection (right panels of \Cref{xz_radec}). In this projection, the slightly `lower' orbital path is the trailing tail, and the `higher' orbital path is the leading tail. 

We examine the relative proper motion with respect to the cluster to determine the kinematic signature of each tail. Figure \ref{relative} shows the relative proper motion in RA, $\Delta{\mu_{\alpha}}$, (where $\mu_{\alpha}=\mu_{\alpha}\cos\delta$, i.e. true arc proper motion in the direction of right ascension) and in declination, $\Delta{\mu_{\delta}}$ for stars in the simulation that are within a box on the sky of length 13.5 times the tidal radius of the cluster. We color the plot by $\Delta{\mu_{\delta}}$ purely to aid comparison with a later figure. The trailing tail extends upwards from the cluster at $(\Delta\mu_{\alpha},\Delta\mu_{\delta})=(0,0)$ to positive ${\Delta{\mu_{\delta}}}$ and the leading tail extends downwards to negative $\Delta{\mu_{\delta}}$. There is no discernible difference in $\Delta\mu_{\alpha}$ for the majority of the stars in both tails.

\section{Observational data from Gaia DR2}
\label{obs}
To search for potential extra-tidal stars around NGC 288, we first select all stars observed by $Gaia$ within a box of length 13.5 times the tidal radius centred on NGC 288 (where the tidal radius is 31.5 arcmin; \citealt{baumgardt10}). The resulting dataset contains 117,480 stars for which we have astrometric and photometric data with respective errors.

The left panel of \Cref{CMD} shows the colour-magnitude diagram (CMD) for the dataset queried from $Gaia$. Though we cannot directly select cluster stars by their $Gaia$ parallaxes, $\pi$, due to high errors at the distance of NGC 288 (8.9 kpc), we know stars close to our Sun with good parallaxes are not cluster stars. Thus, we remove any star considered nearby to our Sun at an arbitrarily chosen distance of $d \leq 5$ kpc, with fractional parallax errors of less than 10\% ($\sigma_{\pi}/\pi \leq 0.1$). We calculate distance naively as $d=1/\pi$, since precise distances are not required for this cut. We also remove stars with high proper motion errors, of $\sigma_{\mu_{\alpha}}$ or $ \sigma_{\mu_{\delta}} > 1$ mas yr$^{-1}$. In addition, we exclude data that has less than 8 visibility periods and nonzero astrometric noise, suggested by \citet{flags_ref}. Finally, we remove any stars without colour data leaving 23,224 stars in the sample.

To extract member stars of NGC 288 or recently escaped stars, we aim to identify stars with low relative proper motions compared to NGC 288 that populate a clean Main Sequence (MS) in the CMD. We used the Density-Based Spatial Clustering of Applications with Noise (DBSCAN) algorithm \citep{DB_ref} of the {\sc scikit-learn} module \citep{scikit-learn_DBSCAN} to select clusters of stars from NGC 288 candidates in proper motion, colour, and apparent magnitude space. Parallaxes from $Gaia$ were not used, due to high errors in the parallax data at the distance of NGC 288.

DBSCAN is a cluster finding algorithm that searches for regions of high density in a given parameter space, without a pre-assumed number of clusters to find. Since our parameter inputs to DBSCAN include CMD axes, DBSCAN will establish different stellar types as different clusters. The four parameters used in DBSCAN were scaled such that similar ranges were used for both colour and absolute magnitude. This allowed DBSCAN to weight the two properties equally, with a greater weight assigned to the proper motions.

DBSCAN is parameterized by epsilon and minimum samples \citep{scikit-learn_DBSCAN}, which are used to determine cluster membership. DBSCAN begins by assuming all stars in the sample are `noise', not belonging to any cluster. The algorithm steps through each star, determining the number of other stars within an epsilon-radius of the star, where epsilon is measured in this work with a Euclidean distance metric. If the number of stars in the epsilon-radius exceeds minimum samples, the central star is considered a `core star' of the cluster. Once all core stars are identified, DBSCAN classifies any star within the epsilon-radius of a core star that is not itself a core star, as a `border star', leaving the remainder as noise. DBSCAN then chains together core stars that can be connected to each other by stars in a shared epsilon-radius to create the final cluster group. 

Using an epsilon distance of 0.028 and a minimum samples of 23, DBSCAN finds 1,711 stars along the MS. The strict epsilon and minimum samples values were chosen as a balance between maximising the size of the sample, while retaining a clear CMD. However, this means it is possible that less stars are found by the DBSCAN process than may truly exist in the GC and its tails. A range of 0.018-0.037 and 10-50 were found to produce similar results and trends when used for epsilon and minimum samples, respectively. Choices of epsilon and minimum samples are interdependent, and many combinations will provide similar results. Furthermore, an increase in epsilon requires an increase in minimum samples to preserve these results.

To illustrate the stars we have determined as current or recent NGC 288 members, we plot the CMD of our final dataset with application of the DBSCAN algorithm (red) over the dataset following the cuts as previously described (grey), in the right panel of \Cref{CMD}. Similarly to \cite{NGC288}, we trace the cluster along the MS. The horizontal branch stars, though prominent in the original data set, are removed by the astrometric excess noise cut. Mass segregation within the GC causes these horizontal branch stars to be contained within the inner regions of the cluster, which we confirmed by examining their $\alpha$ and $\delta$. Thus, the subsequent analysis of the tail stars is unaffected by their removal. 

\section{Comparison}
\label{Comp}

\begin{figure}
	\centering
    \includegraphics[width=\hsize]{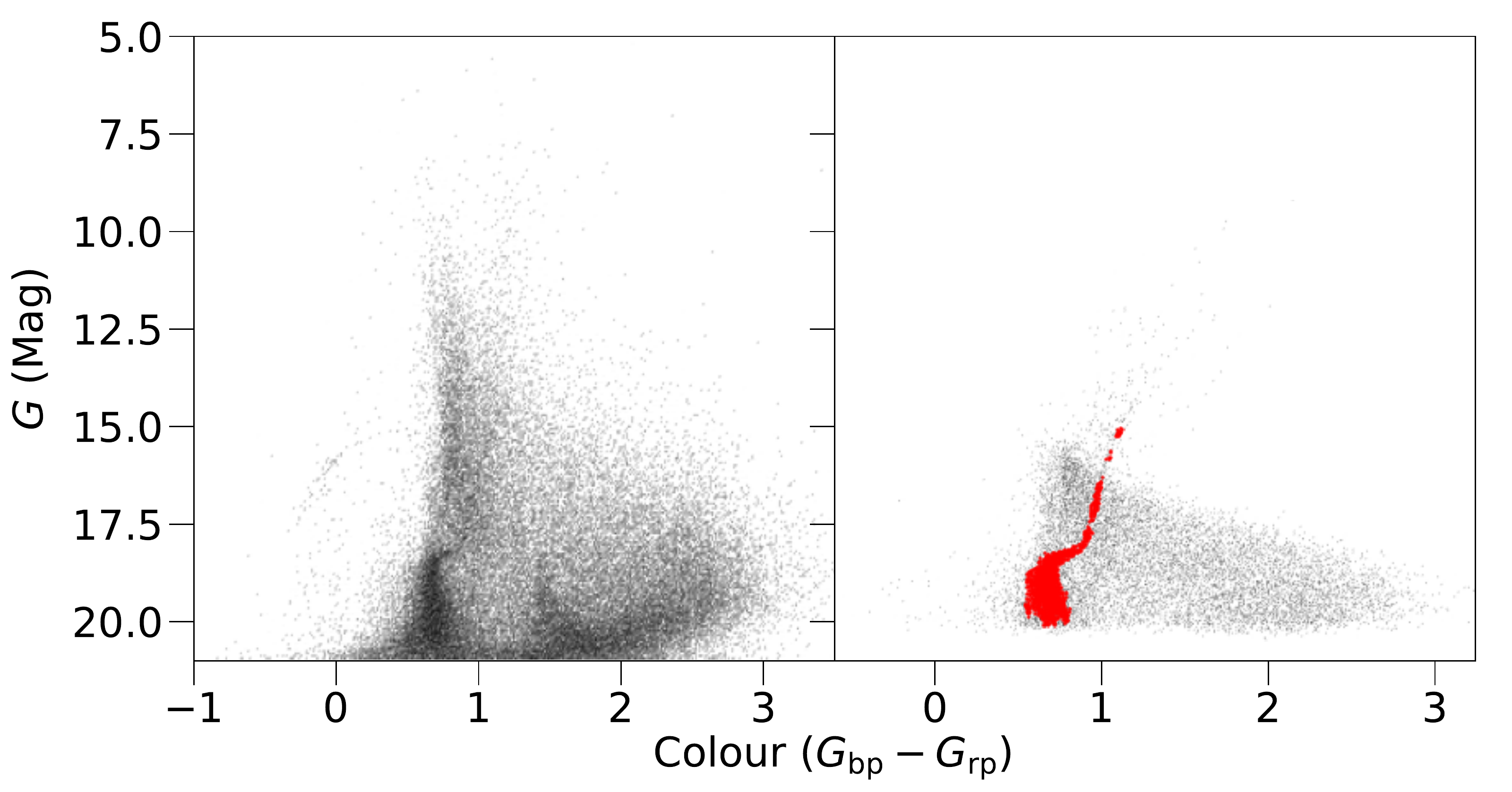}
    \caption{Colour Magnitude Diagram (CMD) for original dataset from $Gaia$ (left panel), results from DBSCAN of a clean CMD (red, right panel), overplotted onto final dataset input for DBSCAN (grey, right panel).}
    \label{CMD}
\end{figure}

\begin{figure}
	\centering
    \includegraphics[width=\hsize]{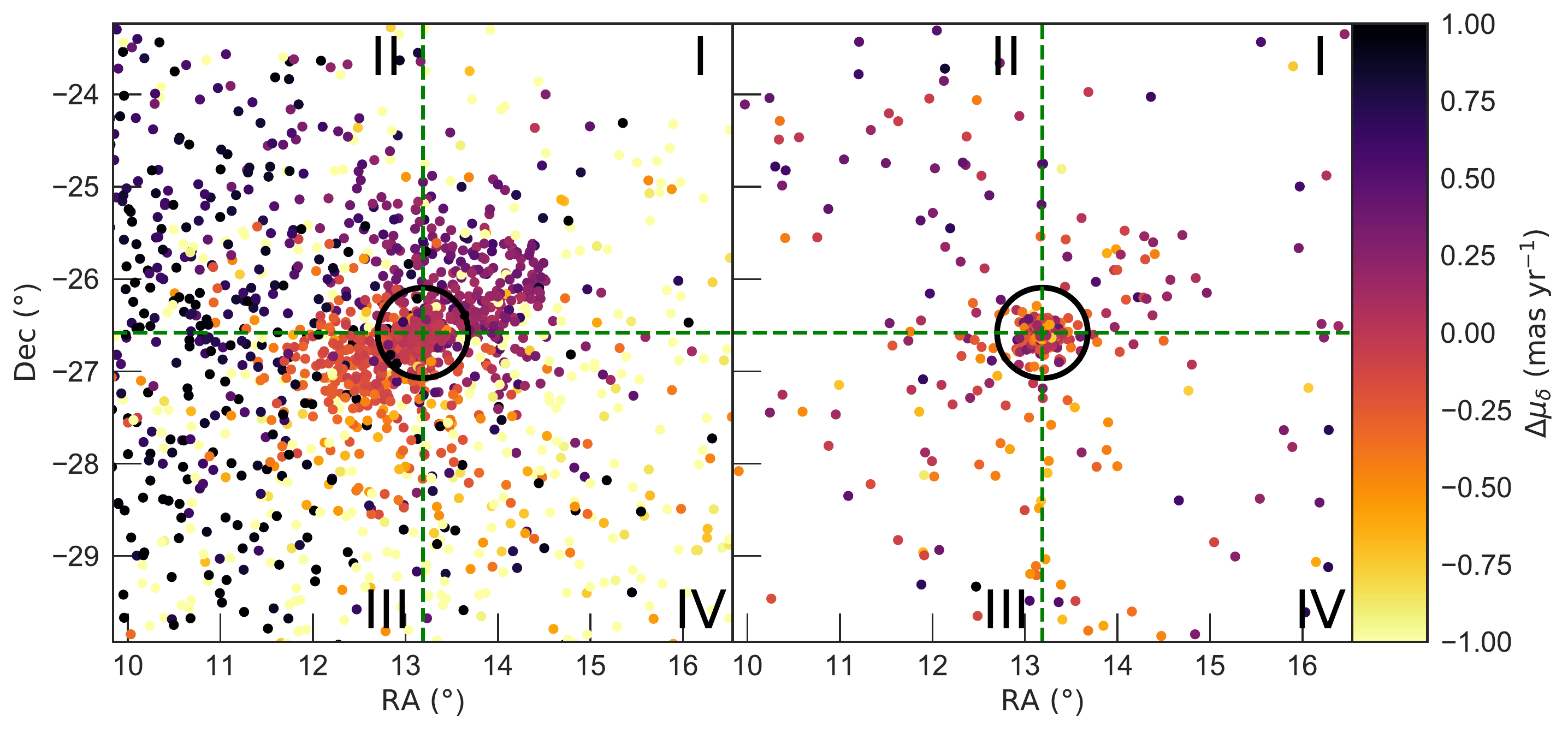}
    \caption{Position of stars in the simulation (left panel) and from the $Gaia$ data (right panel), coloured by proper motion in declination relative to the cluster, $\Delta\mu_{\delta}$. The black circle represents the tidal radius of the cluster, and the dashed lines divide the Figure into four quadrants which are referenced in the analysis. Comparable trends are apparent in the overdensity of stars about the cluster.}
    \label{SimResults}
\end{figure}

Figure \ref{SimResults} shows stars from the simulation (left panel) and the $Gaia$ data after applying the DBSCAN algorithm (right panel) in position-space, coloured by $\Delta\mu_{\delta}$. The black circle represents the tidal radius, demonstrating that many of the DBSCAN selected red points from Figure \ref{CMD} lie outside the tidal radius. We have divided the panels of \Cref{SimResults} into quadrants QI-QIV counterclockwise starting in the top right corner to aid in the analysis.

In both panels of \Cref{SimResults}, a diagonal overdensity of stars is noticeable surrounding the cluster, primarily in QI and QIII. The overdensity in the simulation is composed of tail stars that have recently escaped the cluster. Thus, the fact that we observe the same overdensity in the data implies that stars are escaping NGC 288. A trend in $\Delta\mu_{\delta}$ within the overdensity is clear in the simulation, with model stars in QI predominantly having $\Delta\mu_{\delta} < 0$ and model stars in QIII having $\Delta\mu_{\delta} > 0$. However, it is less clear in the Gaia data, owing to the proper motion uncertainties. The distribution of proper motion error for the tail stars peaks at $\sigma_{\mu_{\delta}}\approx0.2$ mas yr$^{-1}$, with a mean of $\bar{\sigma}_{\mu_{\delta}}\approx0.4$ mas yr$^{-1}$. 

Outside the diagonal overdensity, stars are moving away from the cluster with high relative proper motions. Few stars are seen in QI for both the model and observational clusters. In QII, the majority of stars are moving with high positive proper motion, whereas in QIV a greater number of stars are moving with highly negative proper motion (although some stars do have positive proper motions). In QIII, the distribution of proper motions is reasonably even. These trends are qualitatively consistent between the simulation and the $Gaia$ data.

\begin{figure}
	\centering
    \includegraphics[width=\hsize]{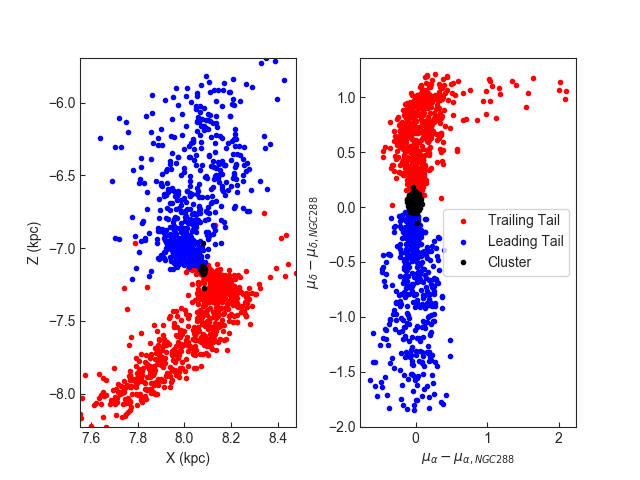}
    \caption{Positions in galactocentric coordinates (left panel) and relative proper motions with respect to the cluster center (right panel) of model stars within a box on the sky of length 13.5 times the tidal radius of the cluster. Stars are colour coded based on whether they are projected within the cluster's tidal radius (black), trailing tidal tail (red), or leading tidal tail (blue).}
    \label{relative}
\end{figure}

To investigate the kinematic signatures seen in both the simulation and Gaia data, we separate model stars within the observed field of view into cluster stars (projected within $r_t$), trailing tail stars, and leading tail stars. This separation is based on star positions in galactocentric coordinates, where the tidal tails are easily identified (see \Cref{xz_radec}). Figure \ref{relative} shows that trailing tail stars (red) have predominantly positive $\Delta\mu_{\delta}$ as they approach apocentre and leading tail stars (blue) have predominantly negative $\Delta\mu_{\delta}$ as they move away from apocentre. 

\begin{figure}
	\centering
    \includegraphics[width=\hsize]{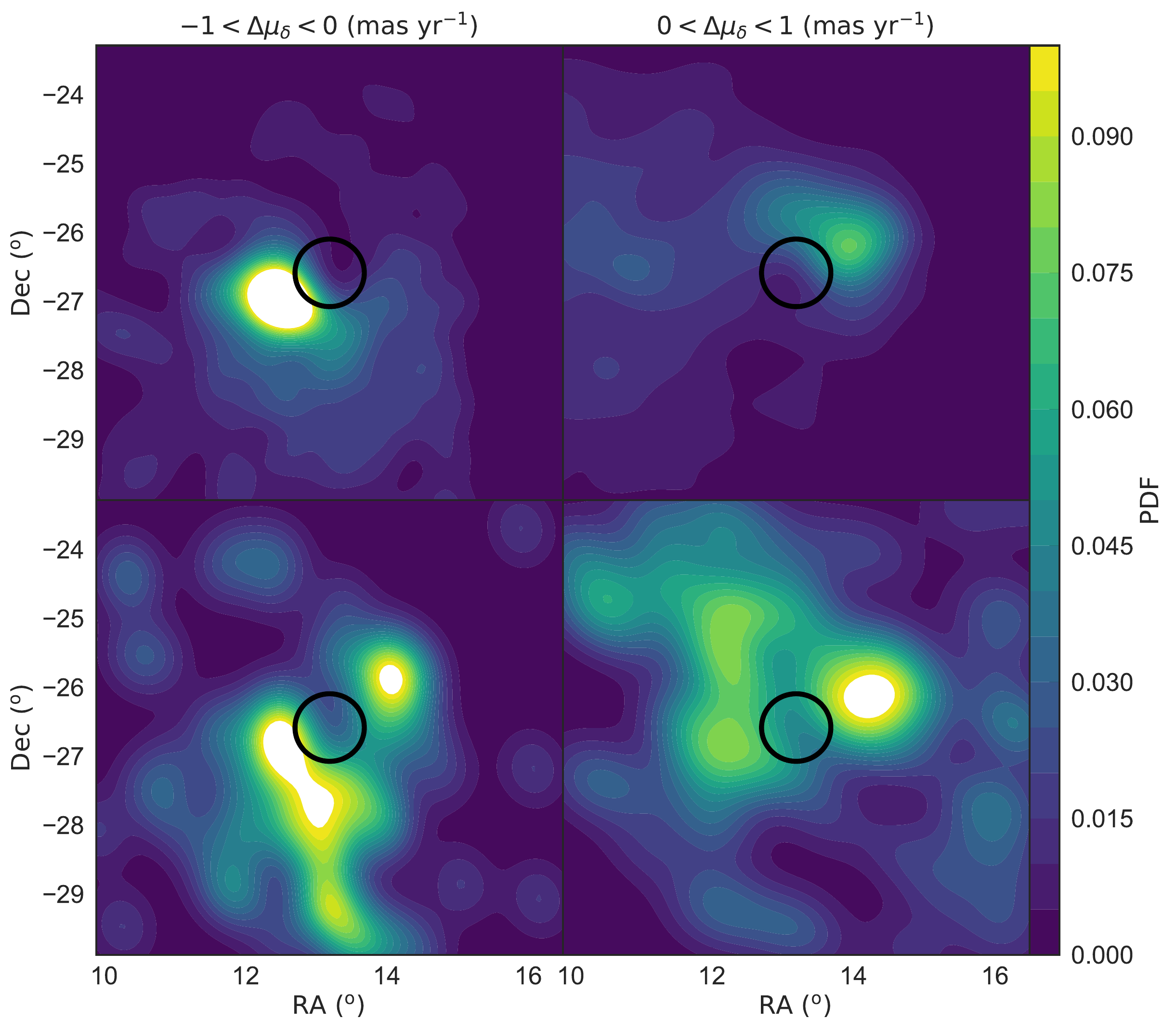}
    \caption{Position of stars outside the tidal radius in both the simulation (top row) and from the $Gaia$ data (bottom row), for stars with $-1<\Delta\mu_{\delta}<0$ mas yr$^{-1}$ (left column) and for stars with $0<\Delta\mu_{\delta}<1$ mas yr$^{-1}$ (right column).}
    \label{contour}
\end{figure}

To better illustrate these trends in the Gaia data, Figure \ref{contour} shows a contour plot for the position of stars outside the tidal radius from the simulation (upper row) and the $Gaia$ data (lower row) split by the sign of $\Delta \mu_{\delta}$. The left column shows stars with $-1<\Delta \mu_{\delta}<0$ mas yr$^{-1}$, which are expected to be leading tail members. The right column shows stars with $0<\Delta \mu_{\delta}<1$ mas yr$^{-1}$, which are expected to be trailing tail members. We keep $\mid\Delta\mu_{\delta}\mid<1$ mas yr$^{-1}$ to select stars which most recently escaped. The distribution of stars in the simulation is dominated by either end of the bar-like overdensity from recently escaped stars. However, the trailing tail is clear in the upper left corner of the top right panel, with the leading tail extending downwards and to the right of the cluster in the top left panel.

The ends of the bar-like overdensity are also clear in the observed data, but the errors are so high that it is not cleanly split between the panels. However, the bar is stronger on the expected side in each panel. The trailing tail with $\Delta \mu_{\delta}>0$ clearly extends towards the upper left corner, and the leading tail with $\Delta \mu_{\delta} < 0$ clearly extends to the left and downwards from the cluster. It is important to note that while qualitatively similar, the exact angle of the tail features is different between the simulations and the observations. This discrepancy is likely due to differences between our assumed potential and the exact tidal field of the Milky Way, which can easily lead to the simulated and observed clusters being slightly out of phase near apocentre. However, the simulation still remains a clear indication of the effects of orbital phase and projection on the sky on tidal tails.

\section{Summary}
\label{Summary}
$Gaia$ DR2 contains photometric and astrometric data for many stars in and around GCs. The newly available proper motions are particularly advantageous in studying GCs, such as NGC 288, where tidal tails are expected to exist, but have yet to be confirmed via observations. We then examine the possibility that while tidal tails do exist around NGC 288, they are difficult to observe due to both the GC’s orbital phase and the observed projection of the GC.

We perform an $N$-Body simulation of a cluster on the same NGC 288 orbit, and find that though the cluster contains tidal tails, projection effects in conjuction with the GC's orbital phase cause the tails to be nearly undetectable at present using stellar positions alone, appearing only as an overdensity around the GC. We compare the simulation with $Gaia$ DR2 observations, and find that the two sets of results display comparable trends in the $\Delta\mu_{\delta}$ of GC stars. We show in both the simulation and observations that stars are moving away from the GC with high proper motions relative to the cluster. In splitting the stars by positive and negative $\Delta\mu_{\delta}$, we resolve the trailing and leading tidal tails on the plane of the sky. High errors in proper motion and low star counts in the observations prevent us from making a rigorous statistical detection, though a good qualitative match between the simulation and the data exists.

Future $Gaia$ releases will have lower uncertainties in proper motions and parallaxes, allowing further investigation into NGC 288's tidal tails. In this work we show NGC 288 is consistent in being a normal tidally filling GC with tidal tails. We do not require a high dark matter content or complex tidal history to reproduce the observations.

\section*{Acknowledgements}
We thank the referee for constructive comments. SK is supported by a Summer Undergraduate Research Project with JH, who is supported by a Dunlap Fellowship both at the Dunlap Institute for Astronomy \& Astrophysics, funded through an endowment established by the Dunlap family, and the University of Toronto. JW \& NPJ acknowledge financial support from the Natural Sciences and Engineering Research Council of Canada (NSERC; funding reference number RGPIN-2015-05235) and Ontario Early Researcher Award (ER16-12-061). RC is supported by NSERC. We thank Tina Peters for assistance. This work has used data from the ESA $Gaia$ mission (\url{http://www.cosmos.esa.int/gaia}), processed by the $Gaia$ Data Processing and Analysis Consortium (DPAC, \url{http://www.cosmos.esa.int/web/gaia/dpac/consortium}). Funding for DPAC has been provided by national institutions, in particular the institutions participating in the $Gaia$ Multilateral Agreement.

\bibliographystyle{mn2e}
\bibliography{ref2}

\begin{thebibliography}{25}
\expandafter\ifx\csname natexlab\endcsname\relax\def\natexlab#1{#1}\fi

\bibitem[{{Arenou} {et~al}\mbox{.}(2018){Arenou}, {Luri}, {Babusiaux},
  {Fabricius}, {Helmi}, {Muraveva}, {Robin}, {Spoto}, {Vallenari}, {Antoja},
  {Cantat-Gaudin}, {Jordi}, {Leclerc}, {Reyl{\'e}}, {Romero-G{\'o}mez}, {Shih},
  {Soria}, {Barache}, {Bossini}, {Bragaglia}, {Breddels}, {Fabrizio},
  {Lambert}, {Marrese}, {Massari}, {Moitinho}, {Robichon}, {Ruiz-Dern},
  {Sordo}, {Veljanoski}, {Eyer}, {Jasniewicz}, {Pancino}, {Soubiran}, {Spagna},
  {Tanga}, {Turon}, \& {Zurbach}}]{flags_ref}
{Arenou} F. {et~al.}, 2018, \aap, 616, A17

\bibitem[{{Baumgardt} {et~al}\mbox{.}(2010{\natexlab{a}}){Baumgardt},
  {Parmentier}, {Gieles}, \& {Vesperini}}]{DM_288}
{Baumgardt} H., {Parmentier} G., {Gieles} M., {Vesperini} E.,
  2010{\natexlab{a}}, \mnras, 401, 1832

\bibitem[{{Baumgardt} {et~al}\mbox{.}(2010{\natexlab{b}}){Baumgardt},
  {Parmentier}, {Gieles}, \& {Vesperini}}]{baumgardt10}
{Baumgardt} H., {Parmentier} G., {Gieles} M., {Vesperini} E.,
  2010{\natexlab{b}}, \mnras, 401, 1832

\bibitem[{{Bovy}(2015)}]{B15}
{Bovy} J., 2015, \apjs, 216, 29

\bibitem[{{Bovy} {et~al}\mbox{.}(2016){Bovy}, {Bahmanyar}, {Fritz}, \&
  {Kallivayalil}}]{Bovy_MWstreams}
{Bovy} J., {Bahmanyar} A., {Fritz} T.~K., {Kallivayalil} N., 2016, \apj, 833,
  31

\bibitem[{{Chambers} {et~al}\mbox{.}(2016){Chambers}, {Magnier}, {Metcalfe},
  {Flewelling}, {Huber}, {Waters}, {Denneau}, {Draper}, {Farrow}, {Finkbeiner},
  {Holmberg}, {Koppenhoefer}, {Price}, {Saglia}, {Schlafly}, {Smartt},
  {Sweeney}, {Wainscoat}, {Burgett}, {Grav}, {Heasley}, {Hodapp}, {Jedicke},
  {Kaiser}, {Kudritzki}, {Luppino}, {Lupton}, {Monet}, {Morgan}, {Onaka},
  {Stubbs}, {Tonry}, {Banados}, {Bell}, {Bender}, {Bernard}, {Botticella},
  {Casertano}, {Chastel}, {Chen}, {Chen}, {Cole}, {Deacon}, {Frenk},
  {Fitzsimmons}, {Gezari}, {Goessl}, {Goggia}, {Goldman}, {Grebel}, {Hambly},
  {Hasinger}, {Heavens}, {Heckman}, {Henderson}, {Henning}, {Holman}, {Hopp},
  {Ip}, {Isani}, {Keyes}, {Koekemoer}, {Kotak}, {Long}, {Lucey}, {Liu},
  {Martin}, {McLean}, {Morganson}, {Murphy}, {Nieto-Santisteban}, {Norberg},
  {Peacock}, {Pier}, {Postman}, {Primak}, {Rae}, {Rest}, {Riess}, {Riffeser},
  {Rix}, {Roser}, {Schilbach}, {Schultz}, {Scolnic}, {Szalay}, {Seitz},
  {Shiao}, {Small}, {Smith}, {Soderblom}, {Taylor}, {Thakar}, {Thiel},
  {Thilker}, {Urata}, {Valenti}, {Walter}, {Watters}, {Werner}, {White},
  {Wood-Vasey}, \& {Wyse}}]{PanSTARRS-1}
{Chambers} K.~C. {et~al.}, 2016, ArXiv e-prints

\bibitem[{{Creasey} {et~al}\mbox{.}(2018){Creasey}, {Sales}, {Peng}, \&
  {Sameie}}]{DMGCs}
{Creasey} P., {Sales} L.~V., {Peng} E.~W., {Sameie} O., 2018, ArXiv e-prints

\bibitem[{{Dehnen}(2000)}]{dehnen00}
{Dehnen} W., 2000, \apjl, 536, L39

\bibitem[{{Dehnen}(2002)}]{dehnen02}
{Dehnen} W., 2002, Journal of Computational Physics, 179, 27

\bibitem[{{DES Collaboration} {et~al}\mbox{.}(2016){DES Collaboration}, Abbott,
  Abdalla, Aleksić, Allam, Amara, Bacon, Balbinot, Banerji, Bechtol,
  Benoit-Lévy, Bernstein, Bertin, Blazek, Bonnett, Bridle, Brooks, Brunner,
  Buckley-Geer, Burke, Caminha, Capozzi, Carlsen, Carnero-Rosell, Carollo,
  Carrasco-Kind, Carretero, Castander, Clerkin, Collett, Conselice, Crocce,
  Cunha, D'Andrea, da~Costa, Davis, Desai, Diehl, Dietrich, Dodelson, Doel,
  Drlica-Wagner, Estrada, Etherington, Evrard, Fabbri, Finley, Flaugher, Foley,
  Fosalba, Frieman, García-Bellido, Gaztanaga, Gerdes, Giannantonio,
  Goldstein, Gruen, Gruendl, Guarnieri, Gutierrez, Hartley, Honscheid, Jain,
  James, Jeltema, Jouvel, Kessler, King, Kirk, Kron, Kuehn, Kuropatkin, Lahav,
  Li, Lima, Lin, Maia, Makler, Manera, Maraston, Marshall, Martini, McMahon,
  Melchior, Merson, Miller, Miquel, Mohr, Morice-Atkinson, Naidoo, Neilsen,
  Nichol, Nord, Ogando, Ostrovski, Palmese, Papadopoulos, Peiris, Peoples,
  Percival, Plazas, Reed, Refregier, Romer, Roodman, Ross, Rozo, Rykoff, Sadeh,
  Sako, Sánchez, Sanchez, Santiago, Scarpine, Schubnell, Sevilla-Noarbe,
  Sheldon, Smith, Smith, Soares-Santos, Sobreira, Soumagnac, Suchyta, Sullivan,
  Swanson, Tarle, Thaler, Thomas, Thomas, Tucker, Vieira, Vikram, Walker,
  Wechsler, Weller, Wester, Whiteway, Wilcox, Yanny, Zhang, \& Zuntz}]{DES2}
{DES Collaboration} {et~al.}, 2016, Monthly Notices of the Royal Astronomical
  Society, 460, 1270

\bibitem[{{Dinescu} {et~al}\mbox{.}(1997){Dinescu}, {Girard}, {van Altena},
  {Mendez}, \& {Lopez}}]{dinescu97}
{Dinescu} D.~I., {Girard} T.~M., {van Altena} W.~F., {Mendez} R.~A., {Lopez}
  C.~E., 1997, \aj, 114, 1014

\bibitem[{{Erkal} {et~al}\mbox{.}(2017){Erkal}, {Koposov}, \&
  {Belokurov}}]{Pal5_2}
{Erkal} D., {Koposov} S.~E., {Belokurov} V., 2017, \mnras, 470, 60

\bibitem[{Ester {et~al}\mbox{.}(1996)Ester, Kriegel, Sander, \& Xu}]{DB_ref}
Ester M., Kriegel H.-P., Sander J., Xu X., 1996, in Proceedings of the Second
  International Conference on Knowledge Discovery and Data Mining, KDD'96, AAAI
  Press, pp. 226--231

\bibitem[{{Gaia Collaboration} {et~al}\mbox{.}(2018{\natexlab{a}}){Gaia
  Collaboration}, {Brown}, {Vallenari}, {Prusti}, {de Bruijne}, {Babusiaux},
  {Bailer-Jones}, {Biermann}, {Evans}, {Eyer}, \& et~al.}]{DR2}
{Gaia Collaboration} {et~al.}, 2018{\natexlab{a}}, \aap, 616, A1

\bibitem[{{Gaia Collaboration} {et~al}\mbox{.}(2018{\natexlab{b}}){Gaia
  Collaboration}, {Helmi}, {van Leeuwen}, {McMillan}, {Massari}, {Antoja},
  {Robin}, {Lindegren}, {Bastian}, {Arenou}, \& et~al.}]{DR2GC}
{Gaia Collaboration} {et~al.}, 2018{\natexlab{b}}, \aap, 616, A12

\bibitem[{{Gaia Collaboration} {et~al}\mbox{.}(2016){Gaia Collaboration},
  {Prusti}, {de Bruijne}, {Brown}, {Vallenari}, {Babusiaux}, {Bailer-Jones},
  {Bastian}, {Biermann}, {Evans}, \& et~al.}]{GaiaMission}
{Gaia Collaboration} {et~al.}, 2016, \aap, 595, A1

\bibitem[{Gnedin {et~al}\mbox{.}(1999)Gnedin, Lee, \& Ostriker}]{tidal_shock1}
Gnedin O.~Y., Lee H.~M., Ostriker J.~P., 1999, The Astrophysical Journal, 522,
  935

\bibitem[{{Harris}(1996)}]{GC_cat}
{Harris} W.~E., 1996, VizieR Online Data Catalog, 7195

\bibitem[{{H{\'e}non}(1961)}]{Henon61}
{H{\'e}non} M., 1961, Annales d'Astrophysique, 24, 369

\bibitem[{Odenkirchen {et~al}\mbox{.}(2003)Odenkirchen, Grebel, Dehnen, Rix,
  Yanny, Newberg, Rockosi, Martínez-Delgado, Brinkmann, \& Pier}]{Pal5_1}
Odenkirchen M. {et~al.}, 2003, The Astronomical Journal, 126, 2385

\bibitem[{Pedregosa {et~al}\mbox{.}(2011)Pedregosa, Varoquaux, Gramfort,
  Michel, Thirion, Grisel, Blondel, Prettenhofer, Weiss, Dubourg, Vanderplas,
  Passos, Cournapeau, Brucher, Perrot, \& Duchesnay}]{scikit-learn_DBSCAN}
Pedregosa F. {et~al.}, 2011, Journal of Machine Learning Research, 12, 2825

\bibitem[{{Piatti}(2018)}]{NGC288}
{Piatti} A.~E., 2018, \mnras, 473, 492

\bibitem[{{Shipp} {et~al}\mbox{.}(2018){Shipp}, {Drlica-Wagner}, {Balbinot},
  {Ferguson}, {Erkal}, {Li}, {Bechtol}, {Belokurov}, {Buncher}, {Carollo},
  {Carrasco Kind}, {Kuehn}, {Marshall}, {Pace}, {Rykoff}, {Sevilla-Noarbe},
  {Sheldon}, {Strigari}, {Vivas}, {Yanny}, {Zenteno}, {Abbott}, {Abdalla},
  {Allam}, {Avila}, {Bertin}, {Brooks}, {Burke}, {Carretero}, {Castander},
  {Cawthon}, {Crocce}, {Cunha}, {D'Andrea}, {da Costa}, {Davis}, {De Vicente},
  {Desai}, {Diehl}, {Doel}, {Evrard}, {Flaugher}, {Fosalba}, {Frieman},
  {Garc{\'{\i}}a-Bellido}, {Gaztanaga}, {Gerdes}, {Gruen}, {Gruendl},
  {Gschwend}, {Gutierrez}, {Hartley}, {Honscheid}, {Hoyle}, {James}, {Johnson},
  {Krause}, {Kuropatkin}, {Lahav}, {Lin}, {Maia}, {March}, {Martini},
  {Menanteau}, {Miller}, {Miquel}, {Nichol}, {Plazas}, {Romer}, {Sako},
  {Sanchez}, {Santiago}, {Scarpine}, {Schindler}, {Schubnell}, {Smith},
  {Smith}, {Sobreira}, {Suchyta}, {Swanson}, {Tarle}, {Thomas}, {Tucker},
  {Walker}, {Wechsler}, \& {DES Collaboration}}]{DM_Shipp}
{Shipp} N. {et~al.}, 2018, \apj, 862, 114

\bibitem[{{Sollima} {et~al}\mbox{.}(2016){Sollima}, {Ferraro}, {Lovisi},
  {Contenta}, {Vesperini}, {Origlia}, {Lapenna}, {Lanzoni}, {Mucciarelli},
  {Dalessandro}, \& {Pallanca}}]{DMNGC288}
{Sollima} A. {et~al.}, 2016, \mnras, 462, 1937

\bibitem[{{Teuben}(1995)}]{teuben95}
{Teuben} P., 1995, in Astronomical Society of the Pacific Conference Series,
  Vol.~77, Astronomical Data Analysis Software and Systems IV, {Shaw} R.~A.,
  {Payne} H.~E., {Hayes} J.~J.~E., eds., p. 398

\end{thebibliography}

\label{lastpage}
\end{document}